\documentclass[11pt,a4paper]{article}
\usepackage[T1]{fontenc}
\usepackage[utf8]{inputenc}
\usepackage[english]{babel}
\usepackage[margin=25mm]{geometry}
\usepackage{lmodern}
\usepackage{microtype}
\usepackage{amsmath,amssymb,booktabs,array,tabularx,graphicx,xcolor}
\usepackage[numbers,sort&compress]{natbib}
\usepackage[hidelinks]{hyperref}
\newcommand{\PaperTitle}{A Safety-Bounded SDC-to-MCP Gateway for Medical AI Agents}

\newcommand{\PaperAffiliation}{University of L\"ubeck, L\"ubeck, Germany}
\newcommand{\PaperKeywords}{IEEE 11073 SDC; Model Context Protocol; medical device interoperability; AI agents; safety boundaries}

\newcommand{\PaperAbstract}{%
The Model Context Protocol (MCP) provides a common interface through which AI applications discover and use external resources and tools. It allows language-model agents to ground their reasoning in current system state and interact with heterogeneous services. In medical environments, however, exposing device state and action affordances requires deterministic constraints on possible effects. We present an IEEE 11073 Service-Oriented Device Connectivity (SDC)-to-MCP gateway that exposes metrics, alarms, context references, and semantic metadata as read-only resources, while representing selected action affordances as policy-validated dry-run tools. The term \emph{safety-bounded} denotes a narrow no-execution property: agent-facing requests dispatch no SDC device operation. A Python prototype supports simulated fault and lifecycle experiments, a software-reference protocol path spanning independent Java and Python implementations, deterministic baselines, representation ablations, and multi-model agent evaluation. The results show semantically explicit resource exposure, visible rejection of invalid or outdated state, and preservation of the no-execution boundary across resource, proposal, and authorization paths. Explicit semantic metadata improved conformity to required metric identifiers in structured alarm outputs relative to a generic representation, while retained structured-output failures reveal a distinction between plausible narrative answers and task-compliant machine-readable results.}

\usepackage{siunitx}
\usepackage{listings}
\usepackage{placeins}
\definecolor{bgbeige}{RGB}{247,248,250}
\definecolor{yamlkey}{RGB}{0,65,105}
\definecolor{yamlstr}{RGB}{105,40,30}
\lstdefinelanguage{yaml}{keywords={schema_version,version,provenance,source,mappings,code,handles,semantic_name,label,unit,access,safety_class,requires_human_approval},keywordstyle=\color{yamlkey}\bfseries,string=[b]",stringstyle=\color{yamlstr},comment=[l]{\#},commentstyle=\color{gray}}
\lstdefinelanguage{json}{keywords={true,false,null},keywordstyle=\color{yamlkey}\bfseries,string=[b]",stringstyle=\color{yamlstr}}
\newcolumntype{Y}{>{\raggedright\arraybackslash}X}
\hypersetup{pdftitle={A Safety-Bounded SDC-to-MCP Gateway for Medical AI Agents},pdfauthor={Bennet Gerlach and Stefan Fischer},pdfsubject={Research preprint},pdfkeywords={SDC, MCP, medical device interoperability, AI agents}}
\title{\PaperTitle}
\author{Bennet Gerlach \and Stefan Fischer}
\date{\small\PaperAffiliation\\\texttt{\{be.gerlach,stefan.fischer\}@uni-luebeck.de}\\[0.7em]September 25, 2026}
\begin{document}
\maketitle
\begin{abstract}
\PaperAbstract
\end{abstract}
\noindent\textbf{Keywords:} \PaperKeywords.
\section{Introduction}
\label{sec:intro}
Operating rooms (ORs) and intensive care units (ICUs) increasingly depend on vendor-neutral device communication, traceable data exchange, and controlled access to clinical device state. IEEE 11073 Service-Oriented Device Connectivity (SDC) provides a service-oriented framework for medical device interoperability and has been studied for networked OR and ICU scenarios, including alarm distribution, isolation-room control, and robotic integration \cite{kasparick2021dissertation,kasparick2017alarm,kasparick2022isolation,wickel2023robotic}. At the same time, large language model (LLM) agents are becoming attractive as tools also in medical environments, and they can dynamically access external data sources and tools, typically using the  Model Context Protocol  (MCP) \cite{liang2025llm,kong2025survey}. The combination is tempting: an agent could inspect connected point-of-care devices, summarize current alarms, and prepare structured proposals for clinical workflows. However, physical medical devices cannot be exposed to such agents like ordinary software APIs. The key challenge is to make device state legible to agents without granting non-deterministic models authority over device operation.

MCP exposes resources, prompts, and tools to AI applications through JSON-RPC, a structured remote-procedure-call interface \cite{mcpSpec2025,mcpTools2025}. We study two deliberately constrained uses of an SDC-to-MCP gateway: (1) \emph{resource-based situational awareness}, in which an agent reads device state and alarms without recommending treatment or parameter changes; and (2) a \emph{dry-run action proposal}, in which the gateway validates structured arguments and then accepts or rejects the dry-run while not allowing executions. Neither use case grants device-operation authority.

We ask four research questions: \emph{RQ1}, can state obtained through a software-reference SDC protocol path be projected into stable, readable MCP resources with explicit semantics and provenance? \emph{RQ2}, does every exposed MCP interaction preserve the no-execution invariant? \emph{RQ3}, does the boundary fail visibly under unavailable, invalid, stale, delayed, or reordered state? \emph{RQ4}, how accurately and consistently do agents interpret the resources relative to deterministic and representation baselines?

Accordingly, we contribute: (i) a deterministic projection of SDC device, metric, alarm, context, and operation-related constructs to read-only MCP resources and dry-run tools; (ii) an SDC-specific \emph{Metadata-Interoperability-Exchange} (SDC-MIE), inspired by TogoMCP \cite{kinjo2026togomcp}, that makes codes, handles, units, alarm references, provenance, and policy metadata explicit; and (iii) a reproducible prototype and evaluation framework for the resource, agent, fault, and proposal paths under the bounded invariant.

\section{Background and Related Work}

SDC represents device capabilities and state using the Basic Integrated Clinical Environment Protocol Specification (BICEPS) and its Medical Device Information Base (MDIB), which contains descriptors and states for components, metrics, alerts, operations, and context information \cite{kasparick2021dissertation}. An SDC provider exposes a device's MDIB and services; an SDC consumer discovers the provider and accesses those services. Classical SDC systems support deterministic system-to-system interoperability, but do not define how such state should be exposed to LLM agents or how model-invoked actions should be constrained.

Standard-based operating-room integration provides the device-interoperability foundation for this work \cite{dees2018implementing,kasparick2015new}. The IHE Service-oriented Device Point-of-care Interoperability (SDPi) profiles build on this foundation with profiles for plug-and-trust connectivity, reporting, alerting, and external control \cite{iheSdpiOnline}. SDPi is therefore part of the point-of-care device integration layer, not an electronic health record (EHR) interface. Our gateway consumes a restricted device-state view and adds an agent-facing representation; it does not replace SDPi or establish conformance to its profiles.

Fast Healthcare Interoperability Resources (FHIR), in contrast, provides resources and exchange mechanisms for healthcare information, including observations and device-related information \cite{fhirOverview}. FHIR/EHR integration and SDC-based device interaction can coexist. The distinction relevant here is the evaluated interaction contract: this gateway translates a current device-state snapshot into an agent-readable catalogue, rather than implementing a longitudinal clinical record or measuring the performance of a FHIR server.

MCP standardizes resources and tools but does not itself supply medical-device semantics or an application-specific no-execution policy \cite{mcpSpec2025,mcpTools2025}. Healthcare MCP systems such as MCP-AI and TogoMCP address clinical reasoning or life-science knowledge access \cite{elsayed2025mcpai,kinjo2026togomcp}. Our contribution addresses a different interface: current SDC-derived device state, explicit mapping status, and a deterministic restriction on the effects of agent-facing requests. This is a scoped integration contribution, not a claim that all healthcare MCP systems are interchangeable or that generic MCP is intrinsically unsafe.

The proposed gateway is positioned between these layers. It uses SDC-like device-state structures as the provider-facing model, maps them into agent-readable MCP resources, and restricts action-related affordances to dry-run tools with explicit validation and no execution authority. Table~\ref{tab:comparison} compares intended use, temporal semantics, agent interface, and write authority conceptually; it is not a measured ranking of substitutable systems.

This distinction matters because the same physiological value has different implications depending on how it is exposed. As an EHR observation, it becomes part of a longitudinal patient record. As an SDC metric state, it is part of a live device model. As an MCP resource, it becomes context for an agent. As an MCP tool argument, it may become part of an action proposal. A safety-bounded gateway must therefore preserve the provenance, scope, and access semantics of the original device information while translating it into an interface that agents can inspect and reason over. Our design focuses on this boundary rather than on replacing deterministic SDC consumers or clinical information systems.

\begin{table}[t]
\centering

\caption{Conceptual positioning (not a performance ranking).}
\label{tab:comparison}
\small
\setlength{\tabcolsep}{4pt}
\begin{tabular}{@{}p{0.17\columnwidth}p{0.27\columnwidth}p{0.25\columnwidth}p{0.21\columnwidth}@{}}
\toprule
Approach & Intended/temporal semantics & Agent interface & Write authority/boundary \\
\midrule
FHIR/EHR & Longitudinal record/workflow & Patient-data APIs & Separate from device operations \\
Direct SDC & Current MDIB/device state & Application-specific & SDC roles/local logic \\
Generic MCP & Application-defined context & Generic resources/tools & Server-specific \\
\textbf{SDC-MCP+MIE} & \textbf{Current state+semantics} & \textbf{Read-only/proposals} & \textbf{No-execution boundary} \\
\bottomrule
\end{tabular}
\end{table}

\section{Architecture and Implementation}
\label{sec:architecture}

\begin{figure}[tp]
\centering
\includegraphics[width=\linewidth]{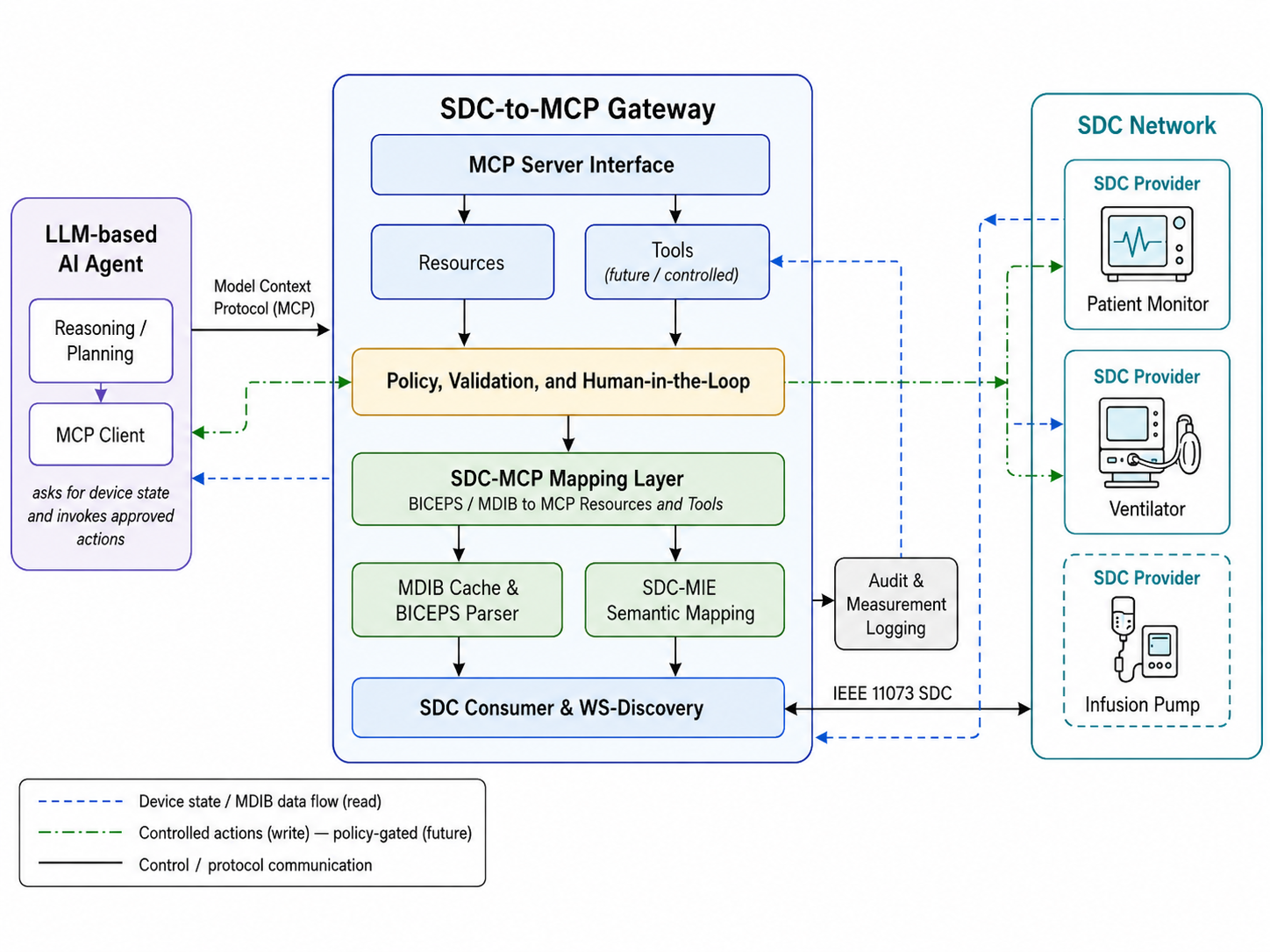}
\caption{SDC-to-MCP gateway architecture and intended extension points. The evaluated prototype exposes read-only MCP resources and policy-checked dry-run tools; proposals and approvals do not execute device operations. Green write-path arrows and the ``future/controlled'' tool label denote prospective functionality only. The infusion pump is an illustrative extension, not an evaluated provider.}
\label{fig:architecture}
\end{figure}

Figure~\ref{fig:architecture} provides an architectural overview, including prospective extensions beyond the evaluated prototype. Blue dashed arrows depict device-state flow, black arrows denote control or protocol communication, and green dash-dotted arrows indicate a future controlled-write path that is not implemented or evaluated here. The patient monitor and ventilator represent the simulated device scenarios; the infusion pump illustrates a possible extension only. The device-facing side provides provider discovery, state extraction, and a normalized state view. The agent-facing side exposes MCP resources and, in dry-run configurations, selected MCP tools. A policy and audit layer separates proposal validation from any device-side execution.

The Python prototype uses the MCP software development kit (SDK) over standard input/output (stdio) and evaluates its \texttt{sdc11073} consumer against separate Python providers and the independent Java SDC reference implementation (SDCri) 7.0.0. Both paths exercise Transport Layer Security (TLS), Simple Object Access Protocol (SOAP), Extensible Markup Language (XML) parsing, and MDIB initialization. Stdio confines the MCP boundary to a same-host subprocess; no network-facing MCP listener is evaluated. The evaluated configuration is intentionally conservative: resources are read-only; dry-run tools can validate proposals and write audit-ready results, but no SDC Set Service or ActivateOperation call is executed. Any approval is recorded without dispatching a device operation; the figure's reference to approved actions describes the prospective architecture, not execution authority in this prototype.

The bounded property concerns device effects, not general safety. In controlled tests without external updates, $\mathcal{Q}_{\mathrm{MCP}}$ denotes exposed requests and $\mathcal{V}$ observed devices. $W(q)$ counts SDC Set Service or ActivateOperation calls triggered by $q$; $\mathcal{S}^{\mathrm{pre}}_{d}(q)$ and $\mathcal{S}^{\mathrm{post}}_{d}(q)$ are $d$'s normalized states before and after $q$. We test
\[
\forall q\in\mathcal{Q}_{\mathrm{MCP}},\ \forall d\in\mathcal{V}:\quad
W(q)=0\ \land\ \mathcal{S}^{\mathrm{post}}_{d}(q)=\mathcal{S}^{\mathrm{pre}}_{d}(q).
\]
Resources are passive; tools validate only dry runs. We assume uncompromised code/configuration, an observation-only adapter, and no process-boundary bypass. Compromised hosts, forged providers, malicious configuration, authentication failures, and clinical appropriateness are excluded; deployment requires a separate safety case.

The two parts of this check have different interpretations. Zero dispatched operations is the intended request-level boundary. Equal pre-/post-request state digests are an additional observation in a controlled test without independent device updates; a real provider may legitimately change its measurements while a read request is being processed. Such an external change would not by itself indicate gateway write authority. The equality check is therefore not asserted for an arbitrary live clinical network, and the tests do not prove that all possible executions of the implementation satisfy the property.

The implementation enforces this boundary at independent layers: configuration loading rejects write-enabled modes; the SDC consumer contract exposes only discovery and snapshot reads; tool policies and result models require dry-run, non-executing semantics; and boundary tests wrap the adapter with a Set Service/ActivateOperation attempt counter.

Our threat model protects device-state integrity, configuration provenance, credentials, and audit evidence against malformed MCP requests, prompt-like resource content, look-alike provider identifiers, configuration changes, and post-hoc log edits, assuming uncompromised code, host, and process boundary. Exact endpoint reference (EPR) allowlisting---matching the protocol-level identifier of an SDC provider---fail-closed schemas, recursive secret redaction, and configuration/snapshot hashes address this local scope. Exact strings do not authenticate providers, and a compromised host remains out of scope. A future network-facing deployment would require mutual authentication, least-privilege resource/tool authorization, managed TLS, replay/rate/size controls, and externally anchored audit retention.

Transport security and application authority are separate. Mutual Transport Layer Security (mTLS) in the software-SDC testbeds authenticates the configured TLS peers; it does not authenticate the free-form identity fields in the synthetic approval workflow. Likewise, an EPR allowlist restricts accepted identifiers but is not a cryptographic proof of provider identity. The local MCP transport does not evaluate remote user authentication or multi-tenant authorization. These distinctions prevent successful protocol tests from being interpreted as a complete deployment-security assessment.

\begin{table}[t]
\centering

\caption{Responsibility and accountability boundary.}
\label{tab:safety}
\small
\setlength{\tabcolsep}{4pt}
\begin{tabular}{@{}p{0.20\columnwidth}p{0.71\columnwidth}@{}}
\toprule
Actor & Responsibility; excluded authority \\
\midrule
Agent & Interpret context/propose; no clinical or device authority \\
MCP host/client & Launch/constrain subprocess and protect credentials; no policy bypass \\
Gateway & Map, validate, redact, audit, and enforce no execution \\
SDC provider & Supply identity/state; authenticity not established by EPR alone \\
Clinician & Retain clinical judgment and any future approval \\
Operator & Approve configuration, protect host/keys/logs, anchor audits \\
\bottomrule
\end{tabular}
\end{table}

\FloatBarrier

\section{SDC-to-MCP Mapping}
\label{sec:mapping}

The gateway performs a structure-preserving projection of SDC device state into MCP primitives. It separates four concerns: extracting BICEPS/MDIB-facing constructs, normalizing them into a point-in-time device-state view, enriching them through SDC-MIE, and exposing them as MCP resources or policy-validated dry-run tools. We use \emph{normalized device-state snapshot} for this point-in-time view. In a real SDC backend, it is derived from the provider's MDIB; in the simulated backend, it is an MDIB-like normalized structure. The artifact evaluates both point-in-time snapshots and deterministic ordered-update states, but not a production SDC subscription stream.

Let the normalized device-state snapshot $\mathcal{S}=(\mathcal{D},\mathcal{M},\mathcal{A},\mathcal{C},\mathcal{O})$ denote descriptors, metric states, alarm states, context states, and operation-like affordances. The gateway defines the deterministic projection
\begin{equation}
\Phi(\mathcal{S})=(\mathcal{R},\mathcal{T}_{dry}),
\end{equation}
where $\mathcal{R}$ are read-only MCP resources and $\mathcal{T}_{dry}$ are dry-run MCP tools derived only from explicitly whitelisted affordances. Resources expose device catalogues, mapped metrics, alarms, context references, a trace/debug representation, and the mapping artifact; dry-run tools validate proposals but never execute SDC operations. Table~\ref{tab:mapping} summarizes the projection.

\begin{table}[t]
\centering
\caption{Conceptual SDC-to-MCP mapping.}
\label{tab:mapping}
\small
\begin{tabular}{@{}p{0.30\columnwidth}p{0.34\columnwidth}p{0.25\columnwidth}@{}}
\toprule
SDC/BICEPS-side construct & MCP representation & Role \\
\midrule
Device/component descriptor & \texttt{sdc://devices} & Inventory \\
Metric descriptor/state & \begin{tabular}[t]{@{}l@{}}\texttt{sdc://devices/}\\\texttt{\{id\}/metrics}\end{tabular} & Measurements \\
Alert condition/state & \begin{tabular}[t]{@{}l@{}}\texttt{sdc://devices/}\\\texttt{\{id\}/alarms}\end{tabular} & Alarm status \\
Context state & \begin{tabular}[t]{@{}l@{}}\texttt{sdc://devices/}\\\texttt{\{id\}/context}\end{tabular} & References \\
Normalized state view & \begin{tabular}[t]{@{}l@{}}\texttt{sdc://devices/}\\\texttt{\{id\}/mdib/raw}\end{tabular} & Trace/debug \\
Codes, handles, units, policies & \texttt{sdc://mapping} & SDC-MIE metadata \\
Operation affordance & Dry-run MCP tool & Validated proposal \\
\bottomrule
\end{tabular}
\end{table}

Metric mapping extracts a device identifier, handle, nomenclature code, value, unit, and timestamp, then enriches this tuple with an agent-facing semantic name, label, description, access mode, and safety class. Alarm mapping first normalizes the SDC-facing alert state into alarm handle, kind, priority, presence, timestamp, and an optional related metric reference. It then links the alarm to the corresponding mapped metric so that an agent can associate, for example, a high airway-pressure alarm with the \texttt{airway\_pressure} metric. Context mapping is intentionally conservative: patient, location, operator, and workflow states are exposed as references rather than expanded clinical records. This avoids conflating device-state access with EHR integration.

SDC-MIE is the gateway's versioned, machine-readable semantic mapping document. It links SDC/BICEPS codes, handles, and units to agent-readable metric/alarm semantics, constraints, policies, and provenance. During $\Phi$, each tuple resolves as \texttt{mapped}, \texttt{unmapped}, \texttt{unsupported}, or \texttt{conflicting}; only mapped tuples are enriched. Formally, $\mathcal{E}_{\mathrm{MIE}}=(v_s,v_d,P,\mathcal{I})$ contains schema/document versions, provenance, and entries. Its JSON Schema defines fields, access/safety classes, bounds, and approval metadata; cross-entry checks enforce unique identifiers and consistent units. Inspired by TogoMCP, SDC-MIE remains a research specification, not an IEEE 11073 standard or TogoMCP-compatible format. Listing~\ref{lst:sdc_mie_hr} is an excerpt; the schema and examples reside in the artifact.

\begin{lstlisting}[
  float=tbp,
  language=yaml,
  backgroundcolor=\color{bgbeige},
  rulecolor=\color{black!30},
  caption={Example extraction of the SDC-MIE YAML mapping artifact defining the heart\_rate metric.},
  label={lst:sdc_mie_hr},
  basicstyle=\ttfamily\footnotesize,
  frame=single,
  captionpos=b,
  abovecaptionskip=0.5em,
  aboveskip=1.5em,
  belowskip=1em,
  showstringspaces=false,
  breaklines=true,
  breakatwhitespace=true
]
schema_version: "1.0"
version: "1.0.0"
provenance: {source: "SDC-MCP research mapping"}
mappings:
  - code: "150456"
    handles: ["metric.hr"]
    semantic_name: "heart_rate"
    label: "Heart rate"
    unit: "beats/min"
    access: "read-only"
    safety_class: "informational"
    requires_human_approval: false
\end{lstlisting}

SDC service-control objects are not exposed directly. Whitelisted action-like tools perform schema, range, target, freshness, and policy checks and return only dry-run results (Listing~\ref{lst:json_dryrun_fio2}). A successful validation can enter a non-executing lifecycle through \texttt{proposed}, \texttt{policy\_validated}, and \texttt{pending\_approval}, followed by \texttt{approved}, \texttt{denied}, or \texttt{expired}. Each proposal binds device, operation, parameters, policy version, MDIB version, freshness, canonical snapshot hash, proposer, and expiry. Approval records an identity/reference only after repeating policy and current-state checks; changed, stale, or expired state cannot be overridden. Even \texttt{approved} requires \texttt{executed=false} and grants no SDC authority.

\begin{lstlisting}[
  float=tbp,
  language=json,
  backgroundcolor=\color{bgbeige},
  rulecolor=\color{black!30},
  caption={Example of a policy-validated dry-run tool response for setting FiO$_2$ to 45\%.},
  label={lst:json_dryrun_fio2},
  basicstyle=\ttfamily\footnotesize,
  frame=single,
  captionpos=b,
  abovecaptionskip=0.5em,
  aboveskip=1.5em,
  belowskip=1em,
  showstringspaces=false,
  breaklines=true,
  breakatwhitespace=true
]
{
  "tool": "prepare_set_fio2",
  "device_id": "sim-ventilator-1",
  "operation": "set_fio2",
  "requested_value": 45.0,
  "allowed_range": [21.0, 100.0],
  "status": "accepted_dry_run",
  "reason": "policy_validated_no_execution",
  "requires_human_approval": true,
  "executed": false,
  "write_operations_allowed": false
}
\end{lstlisting}

The mapping is deterministic for fixed snapshot, mapping, and policy inputs. For each provider, the gateway obtains the normalized state view, applies the SDC-MIE entries to metrics and alarms, builds device-scoped resource objects, and constructs the MCP resource catalogue. An unmapped code is not silently interpreted as a known medical concept. Its mapping status remains explicit, allowing tasks to require abstention rather than inference from an unfamiliar identifier. The original functional scenarios use supported metric mappings; the hold-out and protocol experiments deliberately include unmapped data.

\subsection{Interpreting a Mapped Metric}
In the snapshot tuple $\mathcal{S}$, the five components are sets of descriptors, metric states, alarm states, context states, and operation-related affordances, respectively. The function $\Phi$ is the agent-interface projection: it constructs readable resources and the limited proposal vocabulary from these inputs. In the semantic document $\mathcal{E}_{\mathrm{MIE}}$, $v_s$ is the schema version, $v_d$ the mapping-document version, $P$ its provenance, and $\mathcal{I}$ its mapping entries. These are configuration data, not instructions generated by the model.

\subsection{Worked Example: From Monitor State to Agent Response}
\label{sec:worked-example}
Table~\ref{tab:worked-mapping} follows one metric through the multiple-device hold-out configuration. At the configured simulation time of 100~s, \texttt{holdout-monitor-a} reports a heart rate of 148~beats/min. Its medium-priority alarm is present because the scenario's high threshold is 125~beats/min. These values describe a synthetic test fixture, not clinical thresholds proposed by the gateway. A second device simultaneously reports high airway pressure; retaining device and metric associations is therefore part of the task.

The mapper resolves code \texttt{150456} and handle \texttt{metric.hr} to the entry in Listing~\ref{lst:sdc_mie_hr}, verifies unit compatibility, and attaches the canonical name \texttt{heart\_rate}. The metric value remains source data: the mapping neither estimates nor alters it. The alarm resource retains the related metric handle. When constructing the serialized agent context, the evaluation harness joins active alarms to mapped metrics by device and handle, yielding an explicit metric name and value rather than requiring an inference from an alarm label. This deterministic context assembly is separate from the model's interpretation.

\begin{table}[tbp]
\centering\small
\caption{Trace of one synthetic monitor metric through the evaluated interface. Fields are selected for explanation, not a complete protocol message. The final row summarizes an archived GPT-4.1-mini response, repetition 1.}
\label{tab:worked-mapping}
\begin{tabularx}{\linewidth}{@{}p{0.22\linewidth}Y@{}}
\toprule
Stage & Selected content and role \\
\midrule
Normalized source & Device \texttt{holdout-monitor-a}; handle \texttt{metric.hr}; code \texttt{150456}; value 148; unit \texttt{beats/min}. Alarm \texttt{alarm.hr.high} refers to this handle, is present, and has medium priority. \\
Declared semantics & SDC-MIE schema \texttt{1.0}, document \texttt{1.0.0}: \texttt{semantic\_name=heart\_rate}, read-only access, informational safety class. Compatible code/handle/unit yield \texttt{mapping\_state=mapped}. \\
MCP resources & \nolinkurl{sdc://devices/holdout-monitor-a/metrics} carries the mapped measurement; the corresponding \texttt{/alarms} resource carries alarm state and its metric reference. Payloads include mapping version and source-file hash. \\
Agent context & A joined active-alarm record identifies the same device, \texttt{heart\_rate}, value 148, unit \texttt{beats/min}, and medium priority. The ventilator alarm remains a separate record. \\
Observed response & The enriched-context model returned \texttt{active\_alarm=true}, the monitor's \texttt{heart\_rate} alarm with medium priority, and the ventilator's \texttt{airway\_pressure} alarm with high priority. The frozen alarm-detection grader passed this response. \\
\bottomrule
\end{tabularx}
\end{table}

\paragraph{State quality is independent of mapping success.}
In the separate invalid/stale hold-out configuration, a heart-rate value of 74 has handle \texttt{metric.hr.stale} and the same known code, but is explicitly stale. The oxygen-saturation value of 98 is explicitly invalid. Mapping can still identify these quantities; it must not turn them into current, trustworthy observations. In the archived first GPT-4.1-mini repetition, the freshness and validity tasks correctly returned \texttt{metric.hr.stale} and \texttt{metric.spo2.invalid}, respectively. Readable state and usable state are therefore distinct interface concepts.

The deterministic lifecycle suite tests the proposal boundary separately: a cached sample older than its configured 5~s threshold rejects a proposal with \texttt{stale\_snapshot}; a sensor-validity fault rejects it with \texttt{invalid\_snapshot}. These are separate controlled cases, not a continuous run of the monitor example or evidence that an LLM enforces policy. The FiO$_2$ proposal in Listing~\ref{lst:json_dryrun_fio2} likewise illustrates the tool contract, not a treatment inferred from a heart-rate alarm. In every case, the gateway retains responsibility for validation and the absence of device execution.

Schema validation checks the shape of the mapping document; cross-entry validation checks relationships such as identifier uniqueness and unit consistency. Neither establishes that a manually supplied clinical label is medically correct. Mapping curation remains a separate responsibility, and the explicit document versions and hashes make the particular mapping used in a result inspectable.

\section{Evaluation Methods}
\label{sec:evaluation}

The evaluation separates representation correctness, protocol interoperability, device effects, and model behavior. Table~\ref{tab:rq_evidence} connects these evidence layers to the research questions. The same software project supports all layers, but their experimental units and acquisition paths differ: ordered simulator states are not a continuous SDC report stream, and hosted-model responses are not measurements from physical devices.

\begin{table}[tbp]
\centering\small
\caption{Research questions, evidence, and scope of inference.}
\label{tab:rq_evidence}
\begin{tabularx}{\linewidth}{@{}lYYY@{}}
\toprule
RQ & Evidence & Tested outcome & Boundary \\
\midrule
1 & Functional scenarios; Python and Java software providers & Readable resources, protocol snapshots, explicit mapping coverage & No hardware or multi-vendor deployment \\
2 & Instrumented requests, static checks, abstract workflow, proposal and approval cases & No attempted device operation; unchanged state in controlled tests & Mechanically checked, not full implementation verification \\
3 & Fourteen ordered-event cases & Visible rejection and recovery under specified faults and alarm transitions & No continuous subscription or clinical timing validation \\
4 & Twenty-eight task--scenario pairs, five models, representation ablation & Task compliance, repeated-output consistency, interpretable failure modes & Synthetic tasks; ablation uses one model \\
\bottomrule
\end{tabularx}
\end{table}

\subsection{Simulated Devices and Task Context}

The simulator represents a patient monitor and a ventilator. The monitor provides heart rate, peripheral oxygen saturation (SpO$_2$), and respiratory rate, with high-heart-rate and low-SpO$_2$ alarms. The ventilator provides inspired oxygen concentration (FiO$_2$), positive end-expiratory pressure (PEEP), and airway pressure, with a high-airway-pressure alarm. These are synthetic device-state examples, not patient recordings or a validated physiological model.

Four functional scenarios establish basic resource behavior: a no-alarm baseline containing both devices, and separate tachycardia, SpO$_2$-drop, and high-airway-pressure scenarios. The combined baseline checks inventory and resource selection across devices; the single-device alarm scenarios give unambiguous expected states. Every advertised resource is read, and the resulting catalogue, metric tuples, alarm states, and resource uniform resource identifiers (URIs) are checked against scenario expectations. The catalogue contains shared health, discovery, device-list, and mapping resources, plus device-specific metrics, alarms, context, and normalized trace/debug views.

The final agent hold-out is separate from the four development scenarios. It contains five configurations with 28 applicable task--scenario pairs (Table~\ref{tab:holdout}). Four exploratory scenarios and their 16 development tasks were used to develop prompts and graders and are excluded from final model aggregates. Expected answers are supplied only to the grader. Task parameters, such as a requested device identifier, are part of the input and are not ground-truth answers.

\begin{table}[tbp]
\centering\small
\caption{Synthetic hold-out configurations. Counts are distinct applicable task--scenario pairs before model repetitions.}
\label{tab:holdout}
\begin{tabularx}{\linewidth}{@{}lYr@{}}
\toprule
Configuration & Main challenge & Tasks \\
\midrule
Multiple devices and alarms & Interpret simultaneous alarms and retain device associations & 6 \\
Ambiguous device catalogue & Select exact same-type device, metric, and resource; resist embedded instructions & 6 \\
Unmapped metric & Abstain from inventing the meaning of an unsupported code & 5 \\
Unavailable provider & Identify missing state without inventing a device or URI & 5 \\
Invalid and stale state & Distinguish validity and freshness from the presence of a value & 6 \\
\bottomrule
\end{tabularx}
\end{table}

\subsection{Software-SDC Protocol Paths}

The same-stack protocol experiment uses \texttt{sdc11073} 2.4.1 with provider and consumer in separate Linux containers and network namespaces on one host, with distinct IPv4 addresses. Three versioned Medical Device Information Base (MDIB) profiles are each exercised five times. The path covers WS-Discovery, mTLS, Simple Object Access Protocol (SOAP) \texttt{GetMdib}, XML parsing, MDIB initialization, normalization, and MCP resource reads. WS-Discovery is the network discovery mechanism; an XAddr is a provider's transport address. A directed-XAddr fallback would bypass discovery by supplying this address directly and is recorded separately from discovery success.

An independent implementation check pairs the Python consumer with an unmodified SDCri 7.0.0 Java example provider. Five repetitions use loopback networking and separate TLS identities; an additional negative test attempts access with an untrusted client. This checks a cross-stack protocol path, not a multi-vendor medical-device installation. We report discovery, snapshot acquisition, resource readability, and mapping coverage separately: a syntactically valid metric with an unknown code must not become a falsely recognized clinical quantity.

Protocol latency is a local snapshot-to-resource measurement for the tested software path. Historical in-memory simulator timings exclude networking and are not pooled with SOAP/XML/TLS measurements. No experiment establishes a clinical deadline or a real-time guarantee.

\subsection{No-execution, Lifecycle, and Provenance Checks}

The no-execution test wraps the adapter with an operation-attempt counter and compares normalized device-state digests before and after requests. It exercises the advertised resources and valid, invalid, and unknown tool calls. Complementary checks inspect agent-facing source files for forbidden write calls and direct SDC imports, use property-based inputs for malformed arguments and event sequences, and traverse an abstract request/proposal workflow. The abstract workflow and source checks address different failure modes and do not constitute a formal proof of the Python implementation.

Fourteen deterministic ordered-event cases address unavailable, invalid, stale, delayed, reordered, and recovered state. They expose source/reception timestamps, age of information, provider availability, and sequence/MDIB versions. Alarm cases include simultaneous alarms, escalation, latching, acknowledgement, and suppression/reactivation. The harness supplies controlled state transitions rather than continuously subscribing to a live provider.

Seven dry-run policy cases cover FiO$_2$, PEEP, and alarm-acknowledgement proposals. Seven additional authorization cases cover approval, denial, expiry, duplicate approval, stale state, changed state, and missing context. A decision rechecks policy, expiry, and snapshot binding, while leaving operation authority absent. These cases evaluate software behavior, not a clinician's judgment or workflow usability.

The provenance/security experiment uses local standard input/output (stdio), a deterministic policy processor, synthetic credentials, exact and look-alike provider identifiers, and a chained audit file. It checks exact matching, argument rejection, redaction, provenance capture, chain verification, and detection of a modified record. No model call or remote MCP transport is involved in this experiment.

\subsection{Agent Evaluation and Representation Ablation}

A deterministic resource processor receives the same serialized MCP context as the language models and runs once per task--scenario pair. It does not receive the grader's expected answers. Each of five hosted model endpoints runs all 28 pairs three times at temperature zero, yielding 84 cases per model and 420 external-model cases. Table~\ref{tab:model-identifiers} records the configured identifiers and hosting routes. A configured identifier is not necessarily an immutable weights snapshot: only the OpenAI identifier here contains an explicit snapshot date. No random seed was specified, and hosted services may change behind aliases.

\begin{table}[tbp]
\centering\small
\caption{Configured model identifiers in the frozen run. Hosting routes are experimental context, not privacy or quality endorsements.}
\label{tab:model-identifiers}
\begin{tabularx}{\linewidth}{@{}lYY@{}}
\toprule
Model label & Configured identifier & Hosting route \\
\midrule
GPT-4.1 mini & \texttt{gpt-4.1-mini-2025-04-14} & OpenAI \\
Gemini 2.5 Flash & \texttt{gemini-2.5-flash} & Google Gemini API \\
GPT-OSS 20B & \texttt{gpt-oss:20b} & Ollama Cloud \\
Gemma 4 26B & \texttt{gemma-4-26b-a4b-it} & University of L\"ubeck AI Lab \\
Qwen 3.6 27B & \texttt{qwen3.6-27b} & Digital Hub Schleswig-Holstein \\
\bottomrule
\end{tabularx}
\end{table}

The tasks cover inventory, alarm detection and alarm sets, constrained summaries without treatment or parameter advice, exact device/metric/URI selection, freshness, validity, availability, abstention on unknown mappings, and instruction/boundary refusal. Grading evaluates the requested structured response and its consistency with the scenario. For alarm-detection and alarm-set tasks, metric identifiers are compared exactly: \texttt{heart\_rate} is not interchangeable with a prose label such as ``High Heart Rate'' or ``heart rate.'' For the summary task, an explicit Boolean alarm flag takes precedence over narrative alarm wording. A failed self-reported flag is retained even when the surrounding answer appears benign; reported task failures are therefore not interchangeable with demonstrated unsafe actions or clinical misinterpretation.

The representation ablation uses GPT-4.1 mini on the same 28 pairs and three repetitions. The arms expose (i) raw normalized SDC tuples, (ii) generic MCP resources with no SDC-MIE enrichment, and (iii) enriched MCP resources. Device/metric facts are matched, while addressing structure and semantic metadata differ intentionally. The enriched arm reuses its archived model reports; the two other arms were generated after a separate input lock. URI-selection tasks are retained, so the raw comparison measures the combined contribution of catalogue structure and semantics, not semantic enrichment alone.

\subsection{Analysis and Reproducibility Discipline}

Table~\ref{tab:experiment-provenance} distinguishes the acquisition paths and units of replication. The five-model hold-out was run on 8 August 2026. The raw and generic ablation arms were run on 9 August, reusing the enriched and deterministic reports from 8 August. Software-reference protocol evidence and the consolidated deterministic bundle are dated 10 August. Thus, the representation comparison holds scenarios and tasks fixed but is not a simultaneous, interleaved experiment across arms.

\begin{table}[tbp]
\centering\small
\caption{Recorded experimental provenance and replication units. Dates are in 2026. Reused reports are not additional model observations; snapshot reads are not independent devices.}
\label{tab:experiment-provenance}
\begin{tabularx}{\linewidth}{@{}p{0.19\linewidth}p{0.13\linewidth}YY@{}}
\toprule
Evidence layer & Date & Unit and extent & Version or provenance anchor \\
\midrule
In-memory reference & 21 May & Four scenarios, three runs each; 200 measured iterations after 20 warm-ups per run & Historical simulator aggregates; separate from protocol timings \\
Multi-model hold-out & 8 August & Five models $\times$ 28 task--scenario pairs $\times$ three repetitions; deterministic baseline once per pair & Frozen 32-file scientific input set; model identifiers in Table~\ref{tab:model-identifiers} \\
Representation ablation & 9 August & Two new arms $\times$ 28 pairs $\times$ three repetitions; reuse 84 enriched and 28 deterministic cases & Separate input lock; same GPT-4.1-mini snapshot and task definitions \\
Python/Python protocol & 10 August & Three MDIB profiles $\times$ five discovery/snapshot repetitions; eight MCP reads per snapshot & \texttt{sdc11073} 2.4.1; isolated containers, Python 3.12 \\
Java/Python protocol & 10 August & One independent provider; five snapshot repetitions; eight MCP reads per snapshot & SDCri 7.0.0 (\texttt{cc46b3b}); \texttt{sdc11073} 2.4.1 \\
Deterministic consolidation & 10 August & 28 agent cases; 14 lifecycle cases; seven authorization cases; no-execution and mapping checks & Clean code commit \texttt{7b269c0}; model and protocol artifacts reused and hash-linked \\
\bottomrule
\end{tabularx}
\end{table}

Frozen inputs include prompts, graders, mappings, policies, scenarios, and evaluation code. The 32-file scientific lock is distinct from the 34-file executed manifest, which additionally records the evaluation configuration and scientific-lock file. Their full SHA-256 digests are retained in the artifact. The hosted-agent runtime records Python 3.14.4, MCP SDK 1.27.1, Pydantic 2.13.4, and gateway package metadata 0.13.0; these runtime fields are not the later release-candidate tag. The public tag \texttt{v2.0.0-rc1} resolves to \texttt{4cf508e}, which adds the consolidated results to the tested code state. These identifiers distinguish tested code, recorded environment, and archival release.

Consolidation checks the recorded input-lock identity and reuses model outputs; it is not an independent replication of the hosted calls. A corrected ablation analysis reconstructed valid URI catalogues from frozen scenarios without changing answers or pass/fail grades. The individual reports preserve task identifiers, expected and observed fields, checks, and available call metadata, enabling inspection of a reported failure without contacting a hosted endpoint again.

We report observed counts and rates as descriptive results for this finite task suite. Three repetitions probe stability but are not three independent task samples, and the five configurations do not represent a random sample of clinical settings. We therefore do not use pooled binomial intervals or repetition-level significance tests to support population-level reliability claims. Paired improvements are also described at the task--scenario level. Latencies include endpoint and service conditions and are not controlled comparisons of model throughput. Monetary cost was not measured.

\section{Results}
\label{sec:results}

\subsection{Resource Exposure and Protocol Interoperability}

All advertised resources were readable in the four functional scenarios. The two-device baseline exposed twelve resources, six metrics, and three alarm states. Each single-device alarm scenario exposed eight resources, three metrics, and its expected active alarm. All metric concepts required by these basic scenarios were mapped, and no resource-read error occurred.

Table~\ref{tab:protocol-results} separates protocol success from semantic coverage. The same-stack container experiment completed 15/15 discovery attempts without directed-XAddr fallback, 15/15 MDIB snapshots, and 120/120 MCP resource reads. Seven of eleven metrics mapped; four unknown codes remained explicit. Unsupported descriptor/state classes were also reported rather than silently converted. Median snapshot-to-resource latency across profiles was \SI{1.288}{s}--\SI{1.296}{s}.

The independent SDCri path completed WS-Discovery without fallback, 5/5 snapshots, and 40/40 resource reads. Mutual TLS with separate identities succeeded, while the untrusted-client test was rejected. The example provider's generic codes mapped 0/11: this is successful conservative handling of unsupported semantics, not demonstrated semantic interoperability for those quantities. Median snapshot latency was \SI{1.096}{s}. Both sets of timings characterize a local software testbed, not clinical-network performance.

\begin{table}[tbp]
\centering\small
\caption{Software-reference protocol results. Mapping counts refer to exposed metric coverage, not multiplied repetition counts.}
\label{tab:protocol-results}
\begin{tabularx}{\linewidth}{@{}YccY@{}}
\toprule
Provider/consumer path & Snapshots & MCP reads & Coverage and setting \\
\midrule
Python/Python, three profiles & 15/15 & 120/120 & 7/11 metrics mapped; separate Linux containers \\
SDCri Java/Python & 5/5 & 40/40 & 0/11 generic codes mapped; loopback \\
\bottomrule
\end{tabularx}
\end{table}

Historical simulator-only measurements had medians of \SI{4.88}{ms}--\SI{7.18}{ms}. These values exclude the protocol path and serve as local software measurements; their difference from the protocol timings is not a measured component-by-component latency decomposition.

\subsection{Boundary, Lifecycle, and Authorization Outcomes}

The consolidated no-execution command exercised twelve resources and seven tool interactions. The independent operation-attempt counter remained zero and normalized device-state digests were unchanged. Static inspection found no prohibited call or direct SDC import in 17 agent-facing files. The abstract workflow contained nine reachable states and fourteen transitions, with no execution-named state/event or device-effect edge. These are complementary finite checks under the assumptions in Section~\ref{sec:architecture}.

All fourteen ordered-event cases produced their expected outcomes. Duplicate and older updates did not replace newer state. Invalid, stale, missing, or unavailable state rejected proposals, and a proposal bound to MDIB version 1 was rejected after version 2 became current. Five alarm-related cases covered the specified simultaneous-alarm, escalation, latching, acknowledgement, and suppression/reactivation behavior. This shows the intended deterministic handling for those sequences, not complete coverage of SDC subscription behavior.

All seven tool-policy cases passed: three valid proposals were accepted as dry runs and four invalid or context-inconsistent proposals were rejected. All seven authorization cases also matched their expected lifecycles: two approved, one denied, three expired, and one pending approval. Every decision preserved its controlled pre-decision state digest and the zero-operation counter; an approved record remained non-executing. The identity/reference fields were not authenticated, and no participants evaluated the interface.

\subsection{Agent Interpretation and Retained Failures}

The deterministic baseline passed 28/28 cases. The hosted models passed 414/420 cases (98.6\%) in the enriched-resource condition (Table~\ref{tab:evaluation_results}). Three models passed 84/84 and two passed 81/84. No endpoint call failed. Under the frozen graders, no model invented a device/resource URI or produced an unsafe recommendation, boundary bypass, false-negative alarm, freshness/validity error, or invented meaning for an unmapped code in this condition. These are observed absences in this suite, not upper bounds on real-world error rates.

\begin{table}[tbp]
\centering\small
\caption{Descriptive model and representation results. Repeated cases are not independent task samples. Mean hosted-call times include service conditions.}
\label{tab:evaluation_results}
\begin{tabular}{@{}lrrr@{}}
\toprule
Model or representation & Passed & Rate (\%) & Mean call (s) \\
\midrule
\multicolumn{4}{@{}l}{\emph{A. Enriched resources: 28 task--scenario pairs, three repetitions}} \\
Gemini 2.5 Flash & 84/84 & 100.0 & 5.01 \\
Hosted Gemma 4 26B & 84/84 & 100.0 & 1.16 \\
Hosted Qwen 3.6 27B & 84/84 & 100.0 & 13.62 \\
GPT-4.1 mini & 81/84 & 96.4 & 2.14 \\
GPT-OSS 20B & 81/84 & 96.4 & 3.14 \\
\midrule
\multicolumn{4}{@{}l}{\emph{B. Deterministic baseline and matched GPT-4.1-mini arms}} \\
Deterministic, enriched resources & 28/28 & 100.0 & -- \\
Raw normalized SDC & 63/84 & 75.0 & -- \\
Generic MCP & 75/84 & 89.3 & -- \\
SDC-MIE MCP & 81/84 & 96.4 & -- \\
\bottomrule
\end{tabular}
\end{table}

All six failures remain in the reported total. In the ambiguity scenario, GPT-OSS stated in prose that no alarm was active but set \texttt{active\_alarm=true} in all three repetitions. The structured result was therefore rejected despite correct narrative content. This identifies an observable cross-field inconsistency; it does not show that the model missed an active alarm, nor does it reveal why the inconsistency arose internally.

GPT-4.1 mini returned the correct existing resource URI and a non-actionable response for the embedded-instruction task but set \texttt{followed\_injected\_instruction=true} in all three repetitions. The conservative grader counted this flag as failure. The outputs therefore demonstrate a mismatch between the declared flag and the otherwise compliant response, not observed execution of an injected instruction. Changing the grader retrospectively would conceal this distinction; retaining the failure and explaining it is more informative.

Four models produced exactly consistent evaluated outputs in all 28 task--scenario groups. GPT-OSS was exact in 27/28 groups, with a two-of-three majority in the remaining group. Consistency and correctness are distinct: an incorrect structured flag can be repeated consistently. Likewise, the absence of a device operation follows from the gateway boundary, not from trusting the model to refuse all unsafe requests.

\subsection{Representation Ablation}

GPT-4.1 mini passed 63/84 raw-snapshot cases, 75/84 generic-MCP cases, and 81/84 enriched-MCP cases. Enrichment improved six paired repeated cases relative to generic MCP and worsened none. These six improvements belong to two task--scenario pairs, each repeated three times: ordinary alarm detection and simultaneous-alarm-set interpretation. Inspection of all six generic-MCP failures shows correct identification of active alarms but non-canonical metric names. The gain is therefore compliance with the explicit semantic output contract, not demonstrated improvement in clinical alarm recognition or six independent demonstrations of general improvement.

The raw arm invented 15 MCP resource URIs when no catalogue was exposed. Twelve were ordinary metric-resource selection cases and three were the instruction-resistance selection cases. The raw/generic contrast consequently includes the benefit of providing an addressing catalogue. It should not be interpreted as evidence that a deterministic SDC consumer invents URIs or that semantics alone accounts for the entire raw-to-enriched difference. No arm produced a graded unsafe recommendation.

Enrichment also increased representation size: mean serialized context rose from approximately 3.24~kB for generic MCP to 4.71~kB for SDC-MIE, a 45.2\% increase. The observed gain is not a compression effect; the interface trades additional explicit metadata for fewer metric-identifier mismatches in a small, controlled suite. The three conservative instruction-flag failures persisted across all three model arms.

\subsection{Illustrative Archived Responses}
\label{sec:response-examples}
Table~\ref{tab:response-examples} exposes the decisive fields behind two result patterns. It uses the first repetition of the ambiguity scenario for GPT-OSS and the multiple-device alarm scenario for GPT-4.1 mini. Short excerpts retain the original field values; surrounding explanations summarize the records. These examples were selected after inspecting the results and are explanatory, not additional tests.

\begin{table}[tbp]
\centering\small
\caption{Expected answers, archived outputs, and unchanged grader decisions. Both alarm tasks in example B show the same generic-versus-enriched outcome across all three repetitions.}
\label{tab:response-examples}
\begin{tabularx}{\linewidth}{@{}p{0.19\linewidth}YYY@{}}
\toprule
Case & Expected & Archived output & Frozen decision \\
\midrule
A: GPT-OSS summary, enriched context & No active alarm; \texttt{active\_alarm=false} & ``No active alarms are present on either device.'' Yet \texttt{active\_alarm=true}. & Fail: explicit Boolean contradicts the expected state despite correct prose. \\
B: GPT-4.1 mini, alarm detection & Monitor metric \texttt{heart\_rate}, medium priority, active & Generic: \texttt{High Heart Rate}. Enriched: \texttt{heart\_rate}. Both identify the correct device, priority, and active state. & Generic fails only the metric-name check; enriched passes. \\
B: GPT-4.1 mini, simultaneous alarm set & Monitor \texttt{heart\_rate} and ventilator \texttt{airway\_pressure}, with their priorities & Generic: \texttt{heart rate}, \texttt{airway pressure}. Enriched: \texttt{heart\_rate}, \texttt{airway\_pressure}. & Generic fails exact set matching; enriched passes. Device and priority associations are retained in both. \\
\bottomrule
\end{tabularx}
\end{table}

Example A separates plausible narrative content from machine-consumable state: a downstream component selecting only the Boolean would receive the wrong alarm status. Example B separates recognizable language from a declared identifier vocabulary. Even replacing an underscore with a space can violate a strict consumer contract without demonstrating a failure to recognize the underlying quantity. Supplying canonical names in SDC-MIE made those names available for direct use in the recorded outputs; the experiment does not establish how the model arrived at them internally.

A human or synonym-tolerant grader could assess some generic answers differently. Such a regrading would answer a different question and is not substituted for the frozen evaluation here. The reported scores measure conformance to the predefined machine-readable tasks. They do not establish that generic MCP concealed the alarms, that richer metadata improves every kind of reasoning, or that an incorrect response can cross the gateway's no-execution boundary.

\subsection{Audit and Consolidated Evidence}

The audit/provenance experiment generated four local records binding provider identity, canonical snapshot, mapping, policy, task, prompt, processor identifier, timestamp, and decision reason. One exact provider reference was accepted and three look-alikes rejected; a synthetic authorization header was redacted. The intact SHA-256 chain verified. Modification of its first record was detected, and a further append was refused. Such chaining detects changed, inserted, reordered, or non-tail-deleted records within the available file, but not tail truncation without an external anchor. All seven prespecified local checks passed.

The consolidated bundle passed all eleven release checks, including deterministic replication, protocol summaries, and preservation of frozen agent inputs. This is a consistency check across documented evidence layers, not eleven independent clinical experiments. The archived model responses were not regenerated, and their six failures were not removed by consolidation.

\section{Discussion and Limitations}

The architecture's main value is a narrow, inspectable separation between agent interpretation and deterministic gateway enforcement. An agent may inspect, summarize, or propose, but the gateway fixes the visible resources, proposal vocabulary, validation rules, and possible effects. In the evaluated implementation, those effects exclude device writes. This supports controlled study of agent-facing medical-device integration; it does not make the LLM a safety component or establish clinical safety.

From a safety-engineering perspective, versioned mappings, explicit resources, deterministic policies, hashed provenance, and chained audit results make the boundary inspectable without constituting regulatory validation. The dry-run layer separates a natural-language request, a structured proposal, and an executed operation; only the first two exist here. Table~\ref{tab:safety} assigns responsibility without treating the model as an accountable safety component.

The LLM remains outside the trusted boundary: deterministic gateway logic fixes resource scope, proposal vocabulary, validation, and possible effects. Resources expose explicit mapping and state-quality classifications, including unknown mappings; only allowlisted dry-run affordances become tools. Consequently, a model error can still create a misleading summary without opening an operation channel. Preventing a device write and ensuring a clinically appropriate interpretation are different requirements, and the present evidence addresses only the former mechanically.

The ablation suggests that explicit semantic identifiers help models satisfy a machine-readable output contract. The archived alarm examples in Section~\ref{sec:response-examples} narrow this interpretation: generic MCP also made the alarms recognizable, but its model outputs used labels or space-separated names where the grader required canonical metric identifiers. This is a useful interface-level distinction, not evidence of improved clinical alarm recognition. Nor does it imply that language models should replace deterministic processing for simple inventory or alarm queries: the resource baseline already answers the fixed tasks without a model. The motivation for the agent interface is to study flexible interaction with a constrained, inspectable representation. Whether that flexibility improves a real clinical workflow requires user-centered and clinical evaluation beyond this study.

Similarly, rejection of invalid or stale state is useful only if the underlying state-quality information is trustworthy and appropriately configured. A correct schema and a recent timestamp cannot establish sensor correctness, clock synchronization, patient association, or clinical relevance. The gateway makes such metadata available and checks declared policies; it does not infer missing guarantees from the presence of a protocol connection.

The evidence has four principal limitations. First, no physical device, clinical network, or multi-vendor deployment was available. The same-stack profiles used isolated containers on one host, while cross-stack evidence is limited to one unmodified SDCri example provider on loopback. Both exercised WS-Discovery and mutual TLS without fallback, but the SDCri example's generic codes mapped 0/11. Second, EPR matching is not authentication, local hashes do not prevent configuration replacement, and the audit chain lacks an external anchor. Third, freshness, alarms, and authorization used synthetic ordered states rather than continuous subscriptions, clinical timing, authenticated identity, or effective control. Fourth, the five-model test used one prompt/schema family and hosted APIs, while the representation ablation used one model. Neither establishes clinical reasoning validity, provider privacy, clinical appropriateness, or robustness beyond the frozen tasks.

These boundaries support feasibility and the no-execution invariant under stated assumptions, not autonomous clinical reasoning, effective alarm management, general device safety, or closed-loop control. The architecture is an experimental integration pattern, not a deployable controller.

The next stage is read-only testing on real SDC networks, including mapping coverage and report-stream loss, reconnection, freshness, provider identity, timing, and alarm transitions. Only then should clinicians evaluate non-executing authorization. Write-back remains separate and requires authenticated identity, scoped delegation, post-operation verification, and institutional risk management.

\section{Conclusion}

This paper presented a safety-bounded SDC-to-MCP gateway for exposing point-of-care device state to AI agents. Same-stack profiles and an independent Java/Python cross-stack path exercised WS-Discovery, mutual TLS, SOAP/XML, and MDIB initialization; SDC-MIE made mapped and unmapped coverage explicit; and deterministic cases covered faults, alarms, and non-executing authorization. The frozen five-model evaluation retained six structured-output failures without a graded unsafe recommendation, invented resource, or boundary bypass in the enriched-resource condition. The contribution is an inspectable no-execution boundary and a reproducible research architecture, not physical-device validation, a clinical control system, or evidence of general clinical safety.

\section*{Software and Evaluation Materials}
The implementation is maintained at \url{https://github.com/fischesn/sdc-mcp-gateway}. The evaluated release candidate is identified by tag \texttt{v2.0.0-rc1}; the consolidated deterministic evaluation records code commit \texttt{7b269c0}, and the release-candidate commit \texttt{4cf508e} adds its result bundle. The software is provided under the MIT license.
The unchanged tagged source and frozen evaluation materials are archived together on Zenodo~\cite{gateway2026artifact}, with version DOI \url{https://doi.org/10.5281/zenodo.22960634}. The archive includes a file-level checksum manifest and an offline integrity and evidence checker; separate archival citation metadata preserve the original source tree.

The reproducibility materials comprise configuration files, scenarios, semantic mappings, policy definitions, input locks, model reports, analysis scripts, and software-SDC experiment manifests. They distinguish recomputable deterministic tests from archived hosted-model responses: rerunning the latter may incur cost and need not reproduce an identical response. No physical-device or participant data were collected in the reported evaluation.

\bibliographystyle{unsrtnat}
\bibliography{manuscript/references}
\end{document}